\documentclass[prb,twocolumn,preprintnumbers,amsmath,amssymb,superscriptaddress]{revtex4}
\usepackage[paperwidth=210mm,paperheight=297mm,centering,hmargin=2.0cm,vmargin=2.6cm]{geometry}
\usepackage{graphicx}
\usepackage{dcolumn}
\usepackage{bm}
\usepackage{upgreek}
\usepackage{amsmath}
\usepackage{url}
\usepackage{hyperref}
\usepackage{lipsum}
\usepackage{inputenc}

\newcommand{\beq}{\begin{equation}}
\newcommand{\eeq}{\end{equation}}
\newcommand{\bea}{\begin{align}}
\newcommand{\eea}{\end{align}}
\newcommand{\beqa}{\begin{eqnarray}}
\newcommand{\eeqa}{\end{eqnarray}}

\begin{document}

\title{Polarization dependent light-matter coupling and highly indistinguishable resonance fluorescence photons from quantum dot-micropillar cavities with elliptical cross-section}

\author{Stefan Gerhardt}
\affiliation{Technische Physik and Wilhelm Conrad R\"ontgen Research Center for Complex Material Systems, Physikalisches Institut,
Universit\"at W\"urzburg, Am Hubland, D-97074 W\"urzburg, Germany}
\author{Michael Deppisch}
\affiliation{Technische Physik and Wilhelm Conrad R\"ontgen Research Center for Complex Material Systems, Physikalisches Institut,
Universit\"at W\"urzburg, Am Hubland, D-97074 W\"urzburg, Germany}
\author{Simon Betzold}
\affiliation{Technische Physik and Wilhelm Conrad R\"ontgen Research Center for Complex Material Systems, Physikalisches Institut,
Universit\"at W\"urzburg, Am Hubland, D-97074 W\"urzburg, Germany}
\author{Tristan H. Harder}
\affiliation{Technische Physik and Wilhelm Conrad R\"ontgen Research Center for Complex Material Systems, Physikalisches Institut,
Universit\"at W\"urzburg, Am Hubland, D-97074 W\"urzburg, Germany}
\author{Timothy C.H. Liew}
\affiliation{Division of Physics and Applied Physics, School of Physical and Mathematical Sciences, Nanyang Technological University, 21 Nanyang Link, Singapore 637371}
\author{Ana Predojevi\'{c}}
\affiliation{Department of Physics, Stockholm University, SE-106 91 Stockholm, Sweden}
\author{Sven H\"ofling}
\affiliation{Technische Physik and Wilhelm Conrad R\"ontgen Research Center for Complex Material Systems, Physikalisches Institut,
Universit\"at W\"urzburg, Am Hubland, D-97074 W\"urzburg, Germany}
\affiliation{SUPA, School of Physics and Astronomy, University of St Andrews, St Andrews, KY16 9SS, United Kingdom}
\author{Christian Schneider}
\affiliation{Technische Physik and Wilhelm Conrad R\"ontgen Research Center for Complex Material Systems, Physikalisches Institut,
Universit\"at W\"urzburg, Am Hubland, D-97074 W\"urzburg, Germany}

\date{\today}

\begin{abstract}

We study the optical properties of coupled quantum dot-microcavity systems with elliptical cross section. First, we develop an analytic model that describes the spectrum of the cavity modes that are split due to the reduced symmetry of the resonator. By coupling the QD emission to the polarized fundamental cavity modes, we observe the vectorial nature of the Purcell enhancement, which depends on the intrinsic polarization of the quantum dot and its relative alignment with respect to the cavity axis. The variable interaction strength of the QD with the polarized cavity modes leads to the observation of strong and weak coupling. Finally, we demonstrate the capability of elliptical micropillars to emit single and highly indistinguishable photons (visibility over 93 $\%$).

\end{abstract}
\maketitle

\section{Introduction}

Compact and high performance on-demand solid state single photon sources are essential building blocks for several applications associated with quantum technologies\, \cite{Pan2012,Kok2007, Brien2007}.  While there is a variety of systems with the capability to emit single photons, \cite{Michler2000, He2015, Kurtsiefer2000, He2017, Santori2007, Koperski2015} the recent development of high performance sources based on InGaAs quantum dots (QDs) in microcavities\, \cite{Uns16_insitu, Ding2016, somaschi2016near} has established a single photon source with unprecedented performance. The main engineering focus in design and fabrication of coupled QD-micropillar devices was set on optimization of the parameters such as single photon purity and indistinguishability, as well as the coupling of light and matter via careful spatial and spectral alignment\, \cite{He2017, dousse2008}. The most striking breakthrough that allowed for outstanding performance of QDs in microcavities as bright sources of indistinguishable photons is associated with pulsed resonant pumping of the system, which simultaneously provides high single photon purities, high photon coherence, and nearly-deterministic excitation of the QD. Nevertheless, this excitation technique is utmost challenging to apply without polarization filtering of the pump laser, even in the presence of an additional spatial filtering\, \cite{hopfmann2016efficient}. As resonantly excited QDs emit a superposition of orthogonal linearly polarized single photons, the total rate of a source which is operated under the cross polarization excitation geometry will be reduced by at least 50 \%. However, a source that by design emits each photon only in a single linear polarization mode, would not be affected by cross-polarization excitation geometry. Such a source would allow for achieving single photons on-demand and with unity efficiency, as required for many quantum technology tasks. In the case of a QD in a micropillar, this implies extraction efficiencies of close to 95 $\%$ and Purcell enhancement factor of 5-10, hence allowing for single photon emission and collection rates in the GHz regime. 
While approaches for active control of the photon polarization in QD-micropillars have been reported\, \cite{unitt2005polarization, daraei2006control, Strauf2007_S4P_polcontrol, Rakher2008polcontrol_stark, reitzenstein2010polarization}, a study devoted to the interplay of dipole anisotropy with the polarization split cavity modes across the regimes of light-matter coupling, as well as the principal capability of such devices to emit single, high purity coherent photons compatible with requirements of quantum technologies, is still missing. Here, we provide such a study, based on high-Q as well as moderate Q-factor elliptical QD-micropillars, and we show that the strong ellipticity that enhances the emission into single linearly polarized modes of a micropillar cavity does not have detrimental effect on the purity and coherence of the emitted photons.   \\
The paper is structured as follows: First, we introduce the technology and design of the investigated samples and we discuss how the structural properties of the elliptical micropillars yield the characteristic mode-spectrum, resulting in controllable, linearly polarized resonances. We provide an analytical model which correctly describes the energy spectrum. Next, we show that the coupling of a single quantum dot emitter to such polarized modes clearly reflects the vectorial character of the light-matter coupling, both in the weak- as well as the strong coupling regime. Finally, we demonstrate that our elliptical micropillar platform is a promising candidate to, exploiting resonant excitation, generate single photons with high indistinguishability.

\section{Experimental details}
\label{experiment}

\begin{figure}[h!]
\begin{center}
\includegraphics[width=0.48\textwidth]{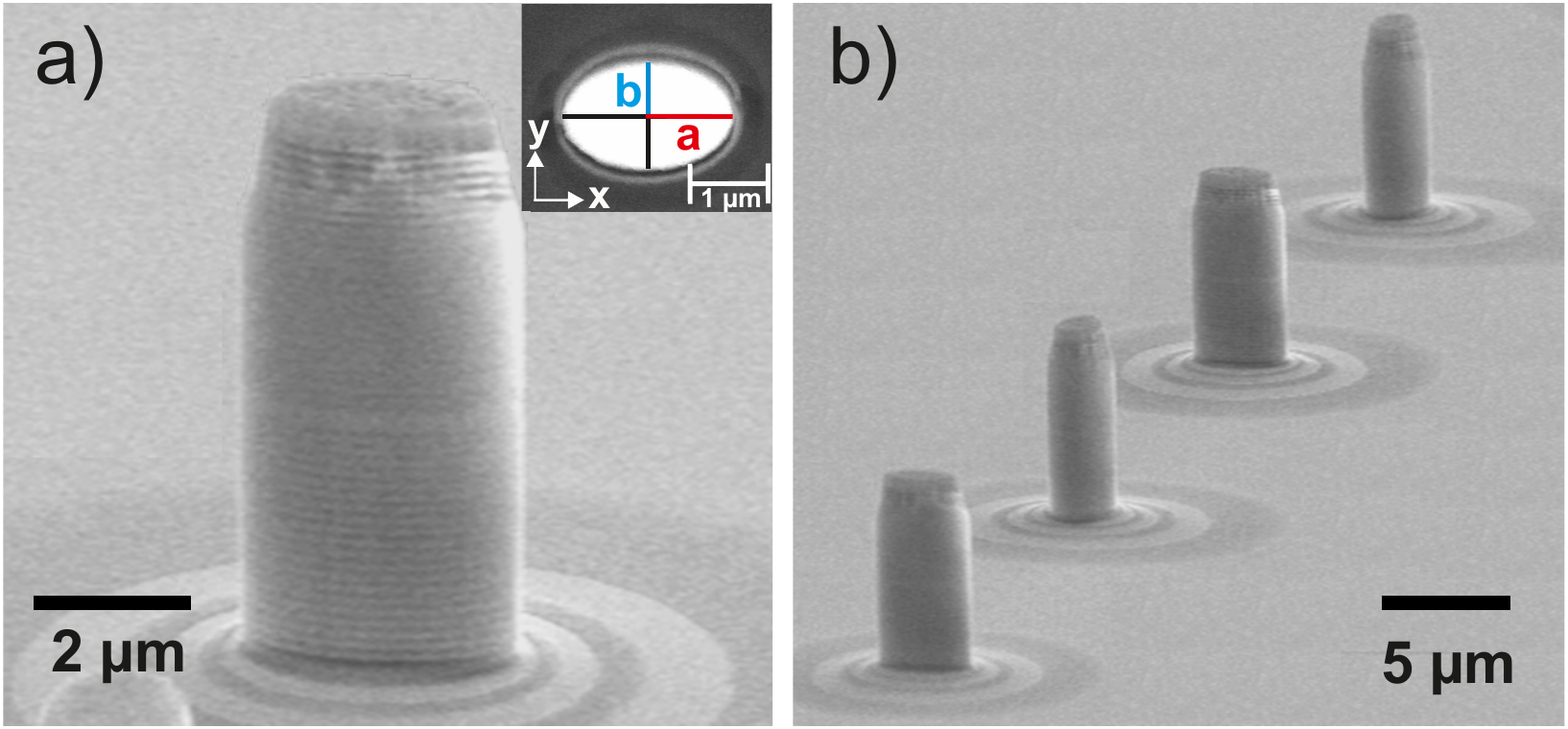}
\caption{(a) SEM-image of a micropillar with an elliptical cross section and an ellipticity of $e$=0.26. The inset displays the top view onto the pillar with the ellipse axes dimensions $a$=2.4\,$\mu$m and $b$=1.5\,$\mu$m. (b) Elliptical micropillars with perpendicular orientations.}
\label{fig:SEM}
\end{center}
\end{figure}

In this work, we studied two samples based on GaAs microcavities grown by molecular beam epitaxy. The first sample contains two stacks of 23 and 27 AlAs/GaAs mirror pairs forming the upper and lower distributed Bragg reflector (DBR). Between the DBR we embedded a $\lambda$-thick GaAs cavity with a layer of low-strain In(Ga)As-QDs as active medium. The indium content was nominally set to 30$\%$. Circular and elliptically shaped micropillars with varied diameters and ellipticities were defined by electron beam lithography and reactive-ion-etching. The scanning electron microscopy (SEM) image in Figure \ref{fig:SEM} (a) depicts an elliptical micropillar. A top view onto the micropillar is plotted in the inset of Figure \ref{fig:SEM} (a). It shows the ellipticity of the structure, with dimensions of 2.4 $\mu$m in x and 1.5 $\mu$m in y direction. The ellipticity $e$ can be calculated as $e = \sqrt{\frac{a}{b}}-1$, in which $a$ and $b$ are the semi-major and semi-minor axis of the ellipse. Figure \ref{fig:SEM} (b) shows several other micropillars with alternating, 90$^\circ$ turned orientations. 
The second sample, which was utilized to generate single, coherent photons described in the last section of this report, was based on a microcavity with 16 and 25 AlAs/GaAs mirror pairs in the top- and bottom DBR, and a single layer of embedded InGaAs QDs grown via the indium flush technique\, \cite{Garcia1998, Wang2006}. After etching the elliptical micropillars, the sample was planarized by Benzocyclobuthene (BCB) and the etch-mask was removed to facilitate resonant spectroscopy on the single-photon level. 

\section{Experimental results and discussion: Elliptical micropillars}
\label{results}

\begin{figure}[h!]
\begin{center}
\includegraphics[width=0.33\textwidth]{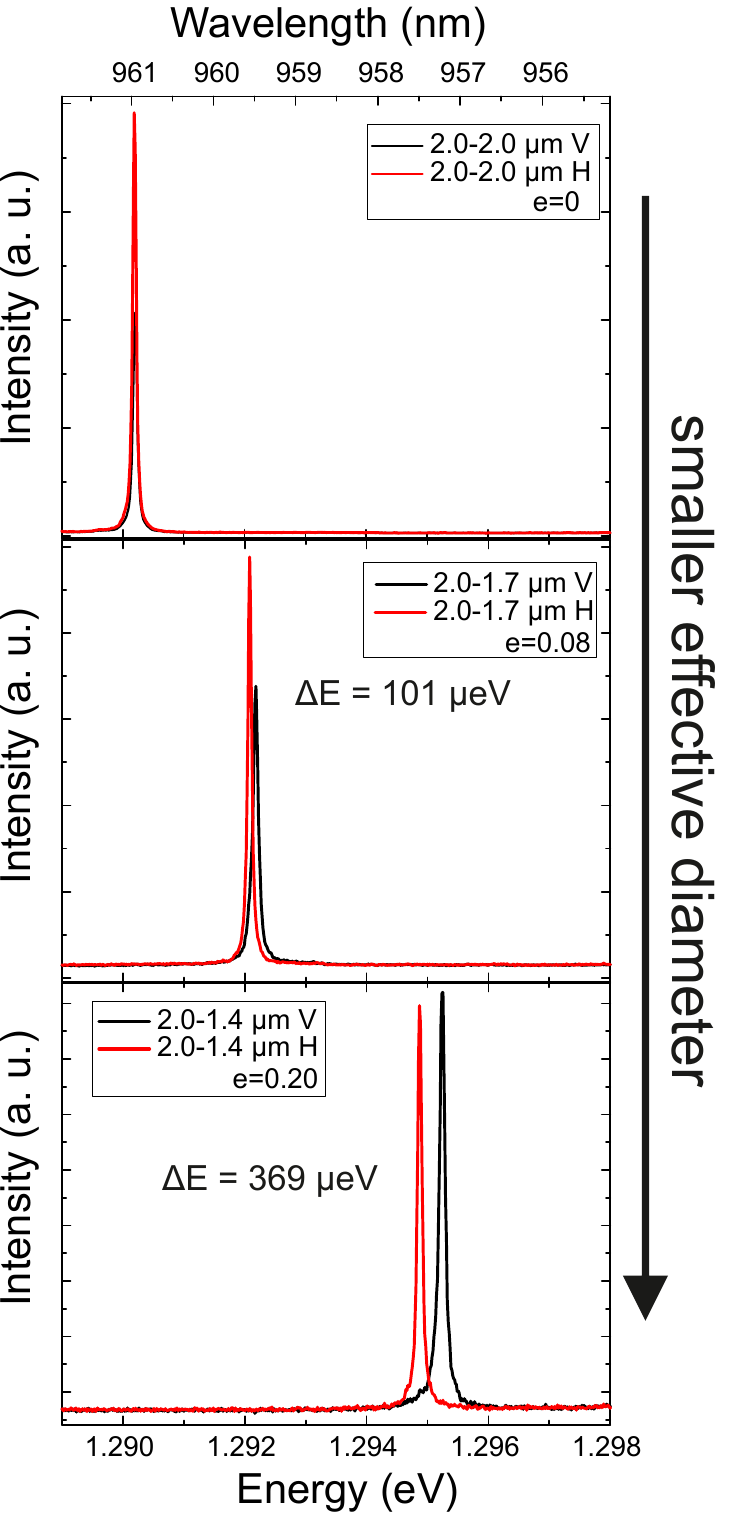}
\caption{Polarization resolved $\mu$PL spectra of micropillars with different value of ellipticity. While circular micropillars do not show a splitting of the fundamental mode, the mode with an ellipticity of $e$=0.2 reveals a splitting of 369\,$\mu$eV. Furthermore, the smaller effective diameter of the micropillar causes a blueshift of the two orthogonal polarized fundamental modes.}
\label{fig:splittingspectra}
\end{center}
\end{figure}

The ellipticity of our micropillar cavities induces a splitting of the fundamental mode, which is typically degenerate for a circular cavity. The two emergent modes support orthogonally linearly polarized light. We study the emergence of the optical resonances in our cavities by non-resonant microphotoluminescence ($\mu$PL) measurements (pump wavelength 532 nm, sample temperature 4.5 K). Exemplary PL spectra of circular and elliptical micropillars are shown in figure \ref{fig:splittingspectra} for pillars with major axis equal to 2 $\mu$m  and minor axis ranging from 2 $\mu$m to  1.4 $\mu$m. These $\mu$PL measurements clearly illustrate the increasing splitting between the two fundamental modes with increased ellipticity as well as the overall blueshift of the resonances induced by the decreasing effective (average) diameter of the micropillars.

\begin{figure}[h!]
\begin{center}
\includegraphics[width=0.42\textwidth]{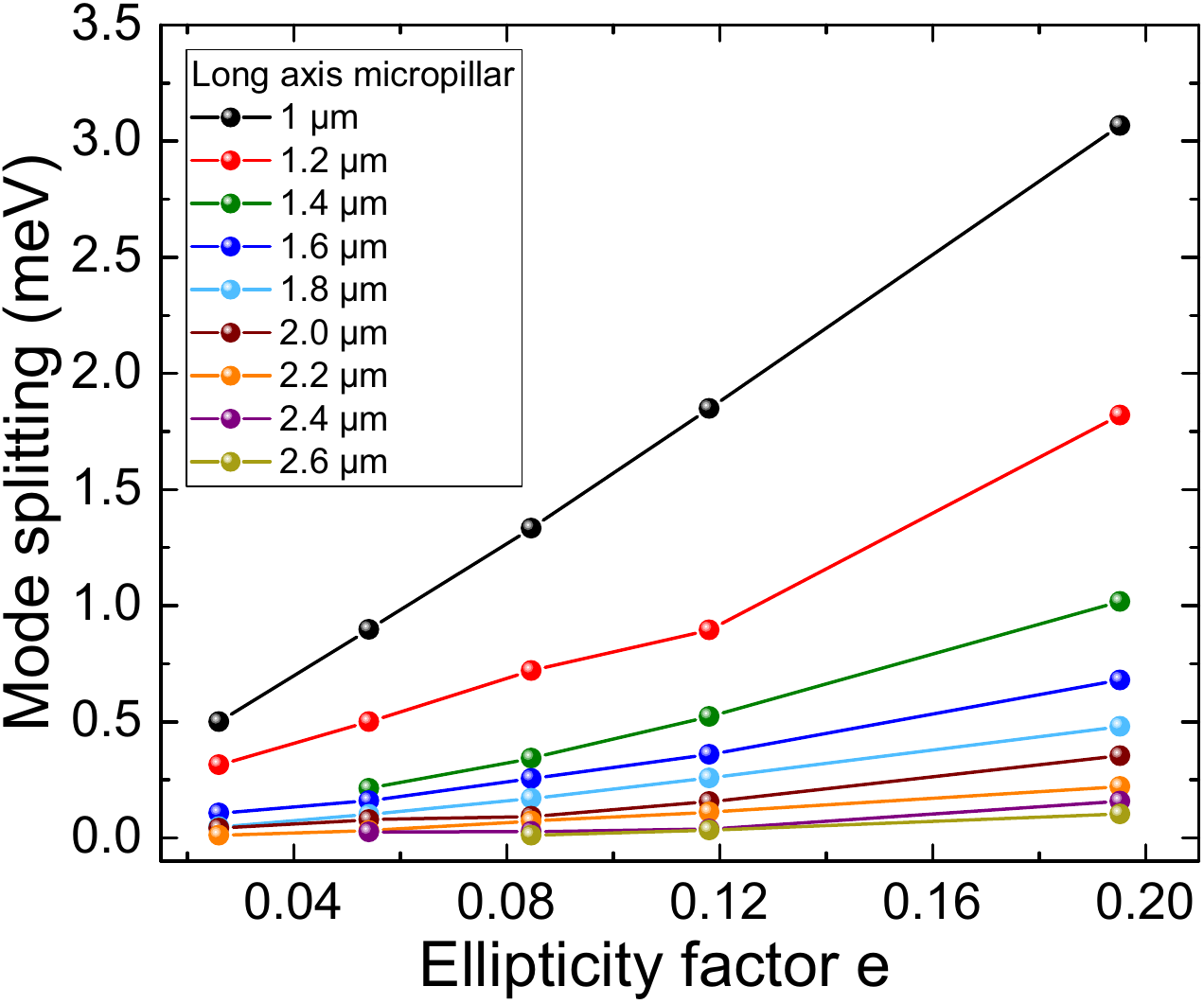}
\caption{Experimentally observed fundamental mode splitting of elliptical micropillars as a function of the ellipticity $e$ with the micropillar major axis (2$a$) as parameter. As can easily be recognized, a smaller dimension of the micropillar leads to a strongly increased splitting of the fundamental mode due to higher lateral confinement.}
\label{fig:modesplitting_ell}
\end{center}
\end{figure}

We extended the study on a number of devices with varied effective diameter and ellipticity. The measured values of splitting of the fundamental mode are shown in Figure \ref{fig:modesplitting_ell}. As foreseen, our measurement also revealed that the mode splitting is strongly sensitive on the micropillar size due to a greater lateral confinement of the photonic modes at smaller diameters. As has already been demonstrated\, \cite{Gayral98}, the mode splitting is dependent on the ellipticity factor $e$. 

\begin{figure}[h!]
\begin{center}
\includegraphics[width=0.48\textwidth]{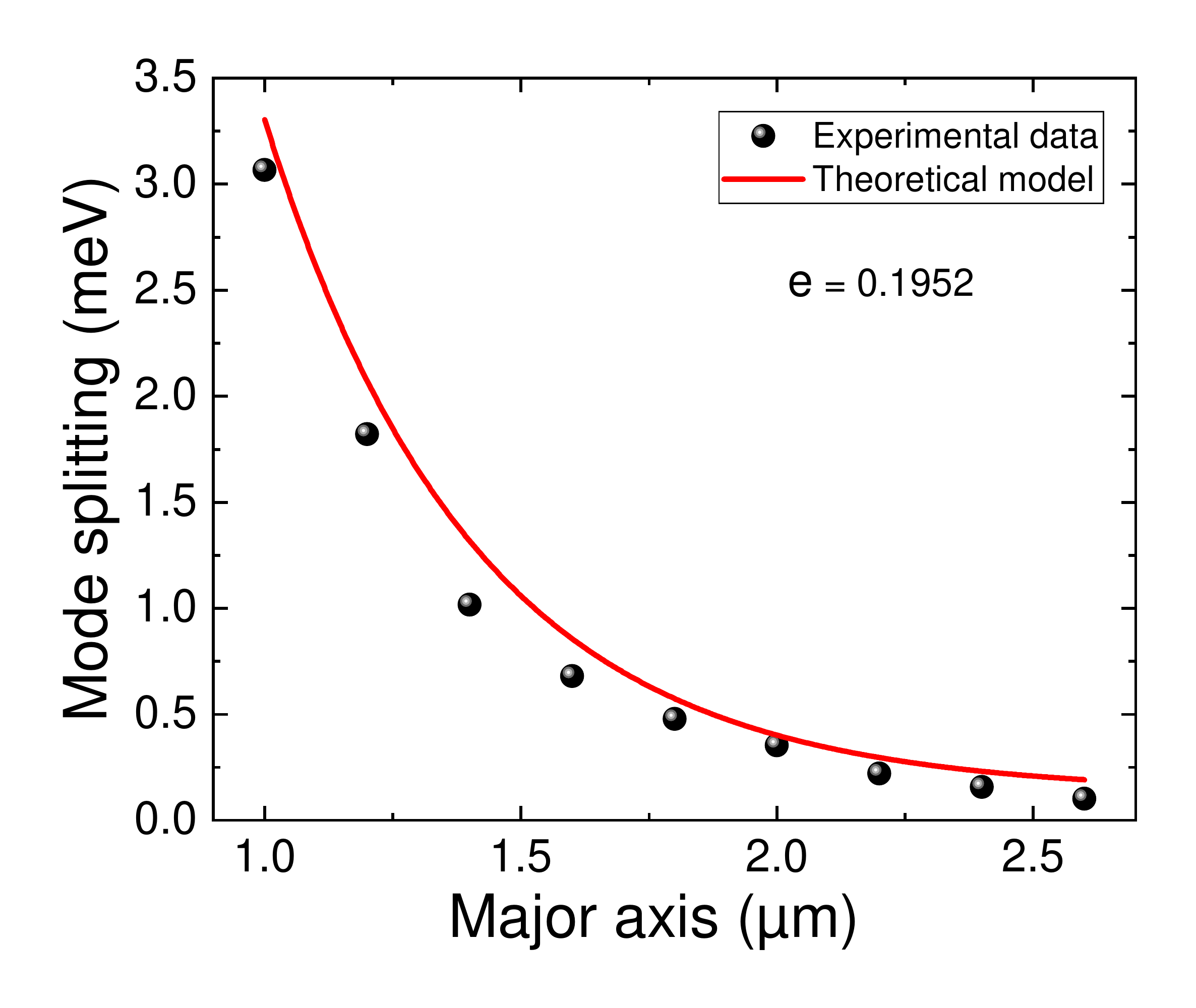}
\caption{Splitting of the fundamental mode for fixed elipticity $e$ as a function of the length of the major axis (2a). The experimental data are in good  agreement with the developed model.}
\label{fig:splitting_theory}
\end{center}
\end{figure}

In Fig. \ref{fig:splitting_theory}, we plot the mode splitting for a series of micropillars with fixed value of ellipticity $e$ = 0.1952 as a function of the length of the major axis of the ellipse. Again, we observe a clear trend towards smaller mode splitting for larger pillars. While in the majority of the previous reports the mode splitting emerging in elliptical pillar cavities has been either analysed numerically\, \cite{daraei2006control} or treated by a phenomenological expression\, \cite{reitzenstein2007ga, reitzenstein2010polarization, sebald2011optical} for the case of a very small ellipticity, here, we establish an analytical model to describe our data. \\

Specifically, we follow the approach of K. Halterman \& P. L. Overfelt\, \cite{Halterman2007}, in which the Maxwell's wave equation is solved approximately for an elliptical waveguide. In contrast to the case of a metallic waveguide, with an exact solution\, \cite{Chu1938}, the solution for dielectrics requires matching of Mathieu functions of the first and the second kind inside and outside the micropillar, respectively, accounting for continuity of different components of the vector field. Unlike the solution using Bessel functions in a cylindrical system, there is no one to one matching of Mathieu functions inside and outside the micropillar as they have different dependencies on the (elliptical) angular coordinate. For this reason, it is necessary to expand the field as a superposition of Mathieu equations and the boundary itself results in coupling between different components. A truncation scheme, limiting to the lowest order Mathieu functions is used to obtain an approximate solution with approximate field matching across the boundary. More details of the calculation are given in the Appendix. The theory, as applied here, is only valid in the limit of small ellipticity. We use the theory here to confirm that the polarization splitting emerges naturally from solving Maxwell's wave equation and that the splitting decreases with the micropillar size. The theory also neglects the significant tapering of the micropillars observed in Fig.~\ref{fig:SEM} and hence is expected to have limited quantitative accuracy. Yet, and despite the notable ellipticity of our system, the corresponding theoretical modelling describes our data very well. The slight overestimation of the splitting could be a consequence of truncating the equation to fulfill the continuity conditions at the microcavity interfaces. However, in contrast to previous approaches, it provides an excellent, and fully transparent approach to analyse the interplay between size, ellipticity and polarization splitting.

\begin{figure*}
\begin{center}
\includegraphics[width=0.96\textwidth]{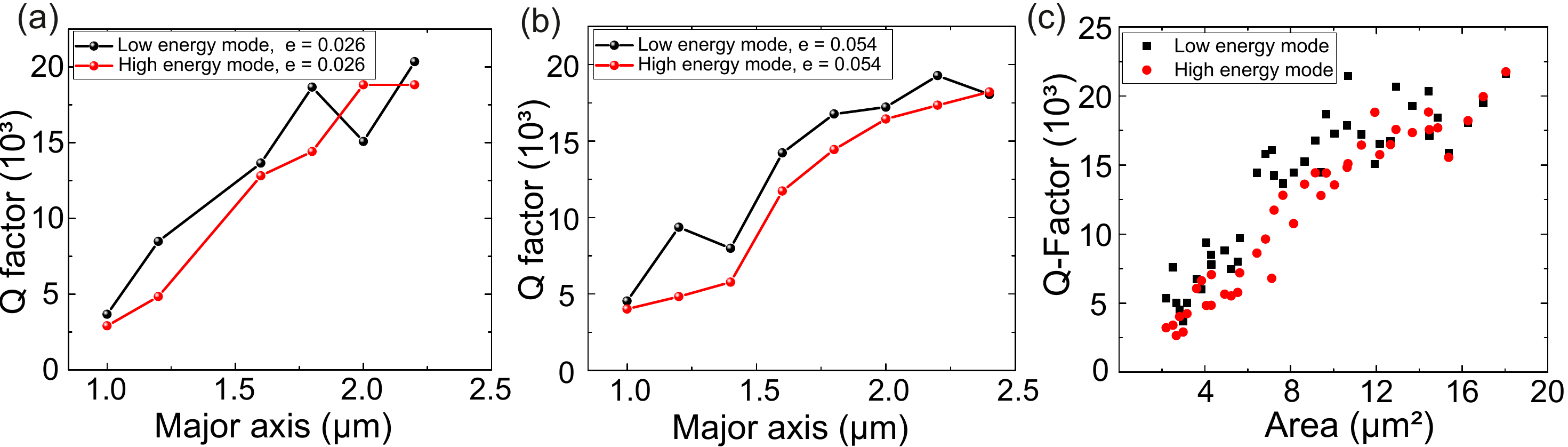}
\caption{(a) Q-factors as a function of the major axis of the elliptical micropillar. A comparison of the two split fundamental modes reveals that the high energy modes of the micropillars systematically yield lower Q-factors.  The same trend can be recognized in (b) with an higher micropillar ellipticity as well as by plotting the Q-factor as a function of the micropillar area in (c). }
\label{fig:Q_factors}
\end{center}
\end{figure*}

Next, we probed the quality of our elliptical micropillar cavities by studying the quality factor $Q = E\,/\,\Delta E$ of the split fundamental cavity modes. Figure \ref{fig:Q_factors} (a) and (b) show the dependency of the Q-factors of the higher and lower energy fundamental modes on the major axis of the elliptical micropillar, plotted for two different ellipticities $e = 0.026$ and $e = 0.054$, respectively. As can be clearly seen, an increasing extension of the micropillar leads to a higher Q-factor due to reduced edge scattering losses, which causes less intrinsic losses\, \cite{Reitzenstein2007}. Here, one recognizes that in general the Q-factor of the high energy mode is lower. This can be explained by a stronger lateral light confinement that opens a loss channel for cavity photons by side-wall scattering\, \cite{Daraei2006, Rei2010}. 

The Q-factors of our elliptical devices rise up to experimentally determined values ranging up to the order of $Q \approx 24 000$, which compares favorably to previously published values\, \cite{whittaker2007high,daraei2007control}. Figure \ref{fig:Q_factors} (c) depicts the Q-factors of the two fundamental modes as a function of the area of the micropillar cavity and supports the observations which were described in \ref{fig:Q_factors} (a) and (b).

\section{Experimental results and discussion: Light-Matter Coupling}

\begin{figure*}
\begin{center}
\includegraphics[width=0.96\textwidth]{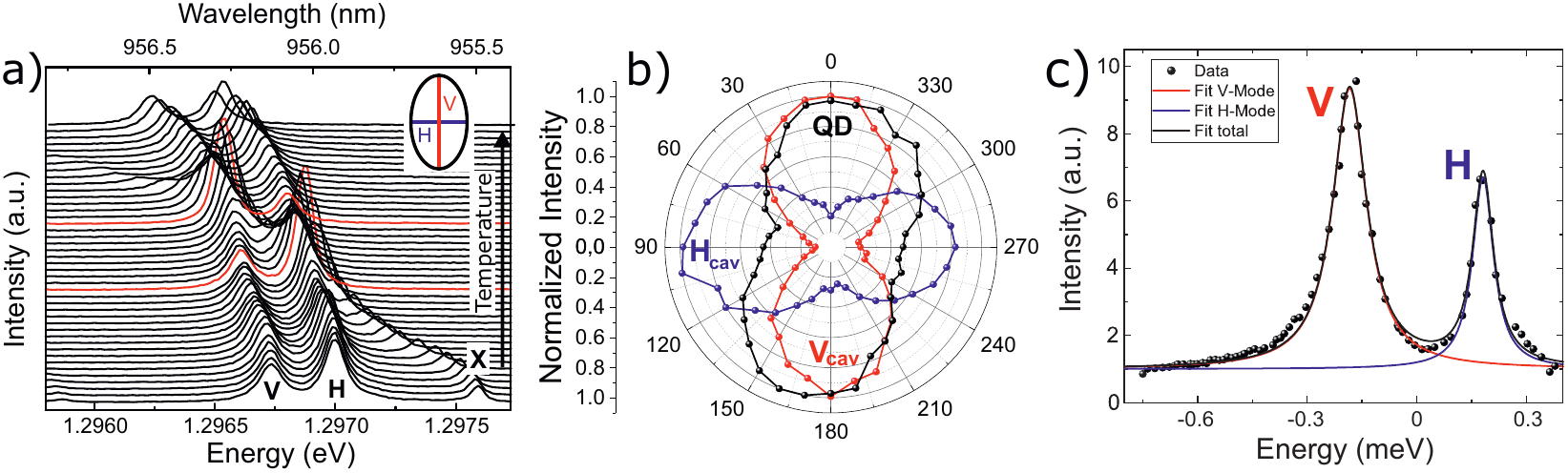}
\caption{(a) Waterfall plot of the temperature series of the recorded spectra, reflecting weak light-matter coupling. The two resonance cases are marked in red. In (b) the polarization of both modes and the QD (off resonant case) is depicted. A high polarization overlap of the QD and the low energy mode can be seen. (c) Plot of the integrated intensity of the QD emission lines as the emitter is tuned through the two resonances. The extracted Purcell factor amount to $F_{P}$ of $F_{P, H}$ = 2.6 $\pm$ 0.7 and $F_{P, V}$ = 6.7 $\pm$ 0.9.}
\label{fig:waterfall_pol_purcell_V}
\end{center}
\end{figure*}

Armed with a detailed understanding of the photonic structure, we turn to the investigation of the coupling of a single quantum emitter to the polarized modes of the elliptical cavities. This study was carried out on a micropillar with 2 $\mu$m long major and 1.4 $\mu$m long minor axis. In the selected device, we identified the luminescence from the neutral exciton of a single QD that at a sample temperature of 4 K occurs at the high energy side of the polarization split cavity resonance. Figure \ref{fig:waterfall_pol_purcell_V}(a) depicts a waterfall plot of a non-resonant $\mu$PL spectra obtained by varying temperature. The upper right black inset indicates the pillar orientation, therefore, $V_{cav}$ and $H_{cav}$ denote the orientation of the vertical and horizontal polarized component of the fundamental cavity-mode while X denotes the QD exciton. With increasing the temperature, the emission of the QD gets red shifted and can be tuned across both polarization modes. For each resonance case, one can clearly observe weak light-matter interaction and a strong enhancement of the emitted QD intensity due to the Purcell effect.

Since the coupling between the QD exciton and the cavity mode should be strongly polarization dependent, we further investigated the linear polarization of our coupled system. In Fig. \ref{fig:waterfall_pol_purcell_V}(b), we plotted the emission intensity of both cavity resonances, as well as the uncoupled QD emission (in the detuned case) as a function of the polarization angle in a polar plot. The two linearly polarized fundamental cavity modes are orientated perpendicular to each other, while the off-resonant QD emission is clearly co-aligned with the low energy mode, here $V_{cav}$. This indicates a strongly anisotropic dipole moment of our QD, which hence should yield a strongly enhanced coupling strength \textbf{d}*\textbf{E} with the co-aligned cavity resonance. Here, we note that the low-strain QDs utilized in this study are subject to a pronounced elongation\, \cite{loffler2006influence}, which explains the strongly directional dipole moment, yielding the pronounced linear polarization in the off-resonant case. Nevertheless, directional anisotropies are inherent to InGaAs QDs grown in the Stranski-Kranstanov method, thus our study is not restricted to this peculiar case.

Since the strong directional anisotropy is expected to yield a polarization dependent coupling strength of our emitter, in Fig. \ref{fig:waterfall_pol_purcell_V}(c) we analyzed the integrated intensity of the QD as a function of the QD-cavity-detuning $\Delta$ with  the particular modes under non-resonant pumping well-below saturation of the QD. To extract the Purcell factor $F_P$ as a measure of the coupling strength, we used the following equation:

\begin{align}
I_{X, cav}(\Delta) \propto \frac{F_P\,L(\Delta)}{1 + F_P\,L(\Delta)} \equiv \beta(\Delta),
\label{eq:I_beta}
\end{align}

where the function $\beta(\Delta)$ quantifies the overlap of the exciton emission pattern with the cavity mode\, \cite{Munsch2009} and $L(\Delta) = 1/(1+\Delta^2/\kappa^2_0)$ is a Lorentzian of width $\kappa_0$ describing the empty cavity line shape.

The fit of the integrated intensity indeed reveals a considerably larger Purcell factor for the resonance case with the lower energy $V_{cav}$ mode compared to the $H_{cav}$ component, namely $F_{P, V}$ = 6.7 $\pm$ 0.9 $>>$ $F_{P, H}$ = 2.9 $\pm$ 0.7. Indeed, the polarization anisotropy of the Purcell-factor reflects the degree of linear polarization of the bare quantum dot on the order of 50-60 $\%$, and thus can be associated with the anisotropy of the oscillator strength of the emitter. We point out, that our result clearly reflects the necessity to engineer not only spatial and spectral properties of QD-cavity systems to optimize light-matter coupling \cite{Uns16_insitu}, but furthermore the polarization properties via precise dipole alignment. 

\begin{figure*}[htb]
\begin{center}
\includegraphics[width=0.99\textwidth]{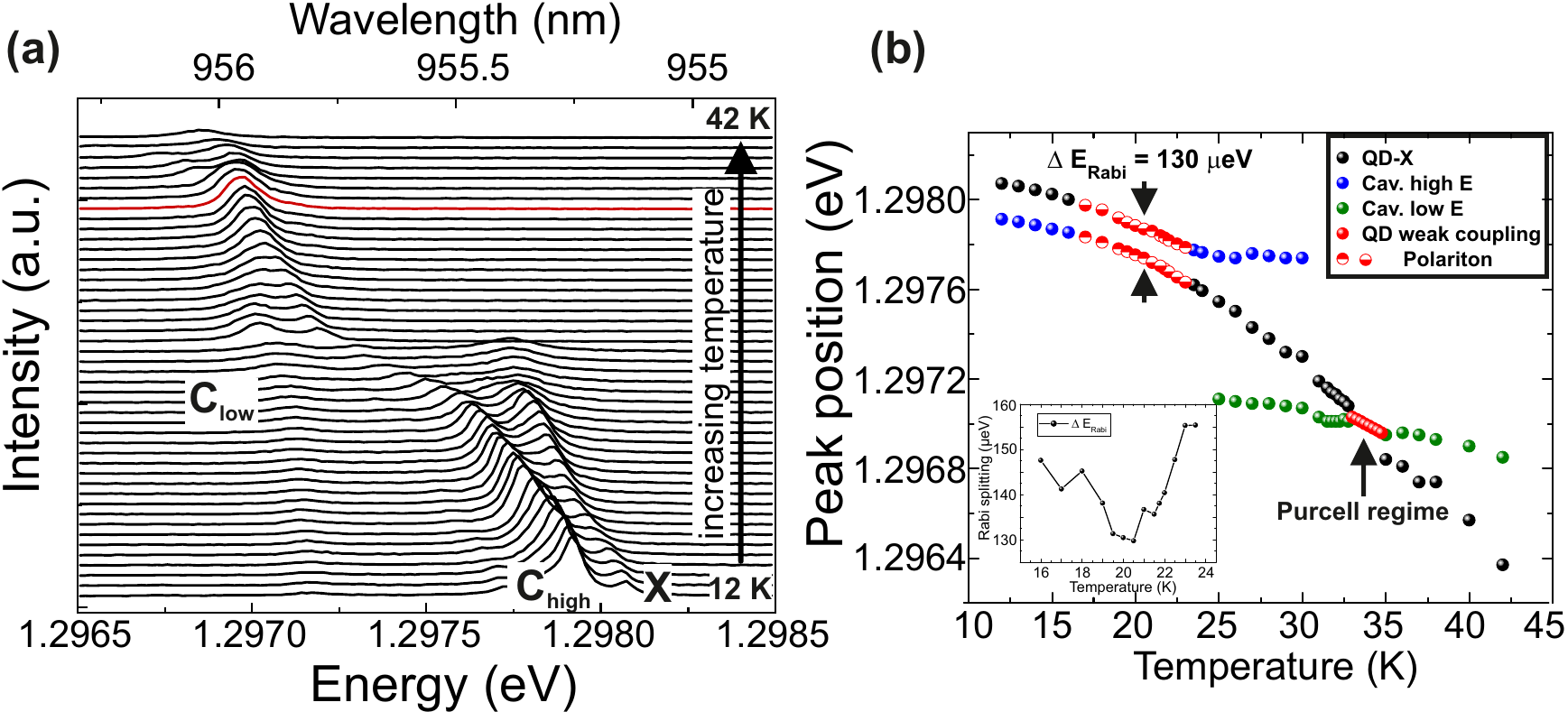}
\caption{(a) Waterfall plot of a temperature series with a QD strongly coupled to the high energy fundamental mode and weakly coupling to the low energy fundamental mode (resonance marked red). (b) shows the peak positions and reveals a Rabi splitting of $\Delta E_{Rabi}$ = 130 $\mu$eV. Again the resonant cases are highlighted in red. The inset depicts the progress of the Rabi splitting with temperature.}
\label{fig:SC}
\end{center}
\end{figure*}

We further conclude that the provided Q-factors and mode volumes of our micropillars put the regime of strong light-matter coupling within reach for selected emitters that are well centered in our devices. Here, we extend our study to a second selected micropillar cavity with such emission features. The cavity is characterized by a major axis of 1.6\,$\mu$m and minor axis of 1.12\,$\mu$m. Fig.\ref{fig:SC}(a) depicts a waterfall plot of non-resonant $\mu$PL spectra obtained by varying the temperature from 12 to 42\,K, where the QD exciton X is tuned into resonance with both the high (C$_{high}$) and low energy  mode (C$_{low}$) of a micropillar. The strong coupling of the QD with the high energy cavity mode C$_{high}$ is evident from a mode anticrossing on resonance. As reported in Ota et al. (2009)\, \cite{Ota2009}, we observe a slight asymmetry in the split-peak spectrum, which we attribute to coupling with acoustic phonons (a linewidth and intensity analysis is provided in the appendix). By further heating up the sample, the exciton shifts through resonance with the low energy mode C$_{low}$, where contrary to the previous resonance a weak enhancement of the emission as the fingerprint of the weak-coupling regime could be observed (resonance red marked). Here, the shift of the QD-exciton X between the two resonances in the waterfall diagram is due to an larger temperature step size taken while recording the spectra.

In order to quantitatively extract the coupling strength of our system, we plot the extracted peak positions in Fig.\ref{fig:SC}(b). With tuning the QD (black dots) into resonance with the high energy mode (red dots) one recognizes an unambiguous anti-crossing, where the two mode branches are separated by the vacuum Rabi splitting. The inset plots the peak separation of the split doublet, yielding a Rabi-splitting as large as $\Delta E_{Rabi}$ = 130 $\mu$eV at 20.5\,K. Further increasing the temperature yields a continuous redshift of the QD until the resonance case is reached with the lower energy cavity mode, where the crossing of the two modes is observed.  

The extracted values for the Rabi-splitting and the linewidth of the high energy fundamental mode and the QD-exciton enables us to estimate the value for the coupling constant $g$ via\, \cite{Andreani1999}:

\begin{align}
g = \sqrt{\left(\frac{\Delta E_{Rabi}}{2}\right)^2 + \frac{(\gamma_{C} - \gamma_{X})^2}{16}}.
\label{eq:g}
\end{align}

This allows us to derive the coupling constant to $g \approx 0.066$ $\mu$eV. Relating this result to the boundary condition to observe strong coupling, which we assess according to Eq. \ref{eq:g} to $g > \frac{\gamma_{C} - \gamma_{X}}{4} \approx 0.017$ $\mu$eV, these estimates explicitly support our observation. Further, the coupling strength can be expressed via the QD's oscillator strength $\textit{f}$ and the cavity's mode volume ${V_M}$ as follows\, \cite{Andreani1999}:

\begin{align}
g = \sqrt{\frac{1}{4 \pi \epsilon_r \epsilon_0} \frac{\pi e^2 f}{m^* V_M}} ,
\label{eq:g_fV}
\end{align}

with m$^*$ being the free electron mass. 
Therefore, by having determined the coupling strength to $g \approx 0.066$ $\mu$eV and approximated the effective mode volume to $V_M \approx$ 0.24 $\mu m^{3}$, we eventually can assess the oscillator strength to $\textit{f}$ $\approx$ 39, confirming previous observations on QDs in microcavities grown via similar techniques\, \cite{Reithmaier2004}.

Although the condition for an observation of strong coupling is also complied for the lower energy cavity mode C$_{low}$, we only observe weak coupling conditions with a very small Purcell enhancement, implying that our QD is simultaneously well-centered in the elliptical micropillar and couples polarization selectively with the two resonances. Since the off-resonant case to study the polarization of the uncoupled QD was not accessible in this particular device, we carried out a statistical analysis of the polarizations of other QDs in the vicinity of the recorded emitter to test our interpretation of polarization anisotropic light-matter coupling: Indeed, we find that the majority of the emitters is strongly linearly polarized, with a main polarization axis being co-aligned with the high-energy cavity mode within an angle of $10 ^{\circ}$. Additionally, previous investigations yielded that for this particular type of QDs the oscillator strength tends to increase with temperatures as a consequence that the dipole in the QD can be thermally activated to overcome tight spatial localization\, \cite{hopfmann2015, Musial2014}.

\section{Elliptical micropillar as a source of single and indistinguishable photons}
\label{results}

Likely the most important application of anisotropic light-matter coupling in QD-microcavities is the generation of highly polarized single photons with high interference contrast, as required in quantum information. While our high-Q microcavity, which has been discussed in the previous sections, is very suitable to observe large Purcell-factors and strong coupling phenomena, a reduction of the Q-factor via reducing the reflectivity of the top DBR is beneficial to increase the photon extraction in single photon sources. 

\begin{figure}[htb]
\begin{center}
\includegraphics[width=0.48\textwidth]{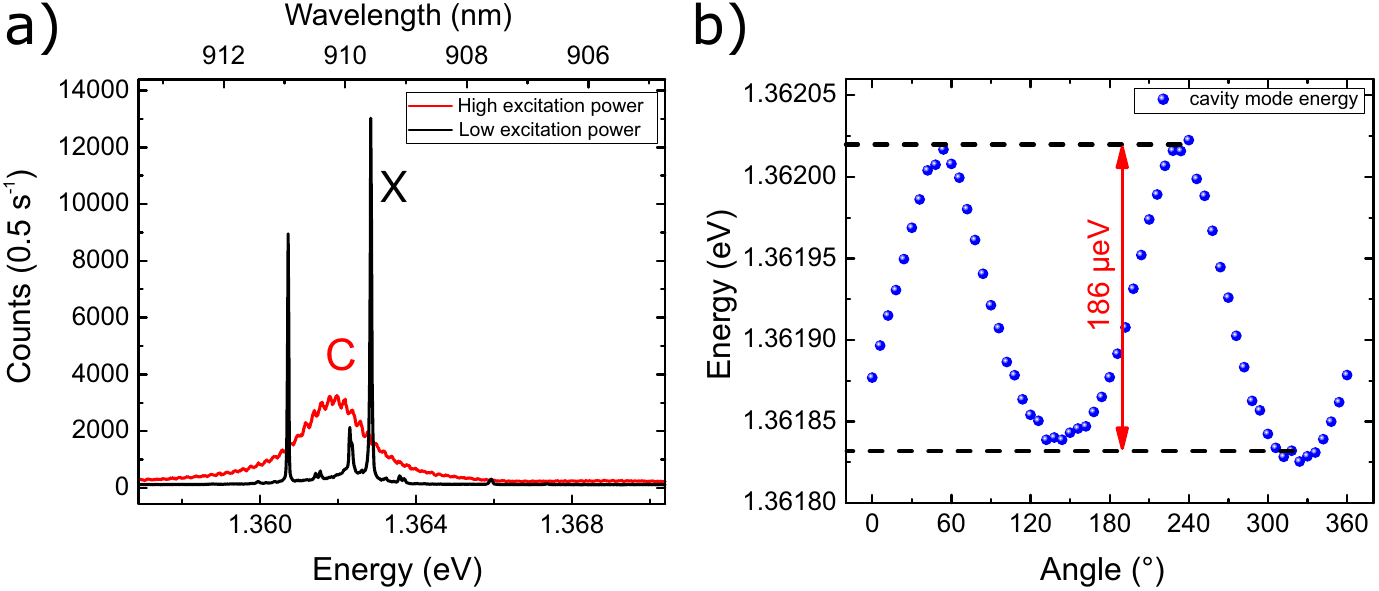}
\caption{(a) Spectra recorded under low and high above-band excitation power. At low excitation power one recognizes several spectrally sharp single emitters, whereas a strong pumping uncovers the broad fundamental cavity mode C of the micropillar. The neutral exciton line of interest has been marked with X. Since the other pronounced line does not show any fine structure splitting, there is a high probability that this line stems from a charged exciton, either of the same, or a neighbouring QD. (b) The polarization-resolved assessment of the spectral position of the fundamental cavity mode reveals a splitting of $\Delta$E $\approx$ 186 $\mu$eV.}
\label{fig:AB_spectra_E_Angle}
\end{center}
\end{figure}

Therefore, for this study we used micropillar cavities with major axis of 3.1 $\mu$m and minor axis of 2.0 $\mu$m, based on the second microcavity wafer which was introduced in section \ref{experiment}. Due to the reduced quality factor, polarization resolved $\mu$PL spectroscopy is necessary to record the splitting of the  fundamental mode of the elliptical micropillar. Fig. \ref{fig:AB_spectra_E_Angle} (a) shows both high and low excitation power spectra, allowing us to assign the spectral position of the cavity modes (C) to the QD line (X) of interest.  

Fig. \ref{fig:AB_spectra_E_Angle} (b) depicts the polarization resolved spectral position of the fundamental cavity mode. The diagram displays a clear sinusoidal behaviour of the energy of the cavity mode, which allows us to extract linear polarization splitting of $\Delta$E $\approx$ 186 $\mu$eV, which is well in line with the ellipticity of $e \approx 0.24$ of the studied micropillar-cavity.  

\begin{figure*}[htb]
\begin{center}
\includegraphics[width=0.98\textwidth]{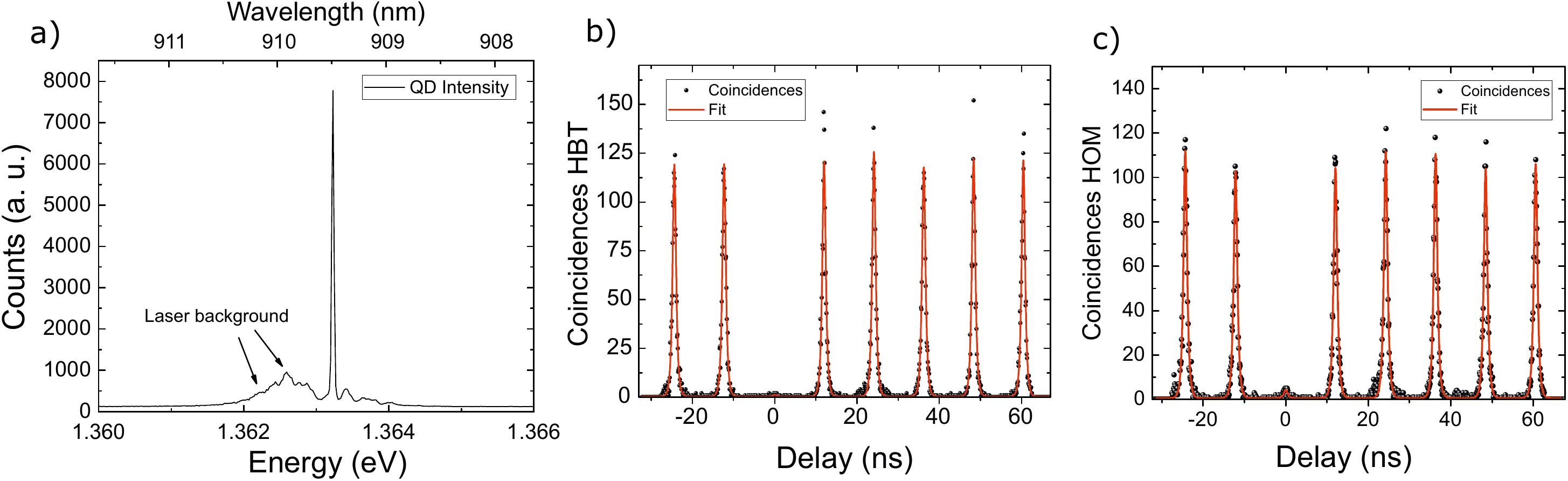}
\caption{(a) Spectrum of the investigated QD using pulsed resonant excitation in a cross-polarization configuration. (b) Second order autocorrelation function recorded under resonant conditions, revealing a multiphoton probability of (0.5 $\pm$ 0.2) \%. (c) Coincidence histogram of the recorded two-photon interference. The clearly suppressed peak around zero delay proves the high degree of indistinguishability of the emitted single photons $\upnu_\textrm{TPI}$ = (93.6 $\pm$ 1.3) \%.}
\label{fig:RF_g2_v2}
\end{center}
\end{figure*}

In order to obtain a high visibility of two-photon interference, it is essential to create single photons of high purity, which here is accomplished by making use of pulsed resonance fluorescence. We applied polarization filtering to suppress scattered light from the excitation laser by approximately seven orders of magnitude. Fig. \ref{fig:RF_g2_v2} (a) depicts the spectrum of the investigated QD, recorded under strictly resonant and pulsed conditions. Beneath the resolution limited spectral emission, one recognizes a weak laser background, which could not be filtered by the cross polarization configuration we used. Nevertheless, a very good straylight suppression in the direct spectral vicinity of the QD is accomplished, which is mandatory for further measurements concerning autocorrelation and coherence properties. For these kind of measurements further suppression of laser straylight was established by spectrally separating the QD signal by means of a monochromator.

The recorded second order autocorrelation function of the QD can be seen in Fig. \ref{fig:RF_g2_v2} (b). The strongly suppressed coincidences around zero delay reflect the purity of the single photon emission from our device. To obtain the $g^{(2)}$(0), we integrated the raw counts in a \textrm{$\pm$}2 ns window around each peak. This yields a $g^{(2)}$(0) = 0.005 $\pm$ 0.002. Fitting the coincidence histogram with a two sided exponential decay convolved with a Gaussian distribution reveals a QD lifetime of $T_1$ $\approx$ 400 ps, which supposes a modest Purcell-enhancement of our QD. \\
The high purity of the single photon emission puts us in the position to test the coherence of the emitted photons from our source by studying the quantum interference of consecutively emitted photons in a Hong-Ou-Mandel experiment, utilizing an unbalanced fiber-based Mach-Zehnder interferometer. 
Analogous to the single photon correlation experiment, the HOM experiment was carried out at a temperature of T = 4.5 K and a close to $\pi$-pulse excitation power of $P_\textrm{Laser}$ = 800 nW. 

 In Fig. \ref{fig:RF_g2_v2} (c), we plot the correlation chart of the two-photon interference experiment. The central correlation peak in our study is strongly suppressed, significantly below the value of 0.5 attainable in the case where the photons would be fully distinguishable. This clearly indicates that quantum interference is established in our experiment. Analogous to the calculation of the $g^{(2)}$(0)-value, we assess the visibility of the interference contrast by comparing the counts in a  \textrm{$\pm$}2 ns window around the central peak for the cases of indistinguishable and distinguishable photons. Latter is set to the half of the average of the seven adjacent peaks, which leads us to a visibility of $\upnu_{TPI}$ = (93.6 $\pm$ 1.3) \%. This value is comparable to previously reported two-photon interferences in high quality quantum dot structures embedded in planar low-Q cavities or circular micropillar structures\, \cite{He2013, Uns16_insitu, He2017, Gerhardt18}. 

\section{Conclusion}
\label{results}

We have carried out a study of single quantum dots embedded in elliptically shaped micropillar cavities. First, we have demonstrated the high quality of our fabrication process, and developed an analytic model to describe the eigenmodes in the system with broken rotation symmetry. We have directly observed polarization dependent strong- and weak light-matter coupling of quantum dot excitons with anisotropic dipole moments in our elliptically shaped pillars. 
In elliptical microcavities with reduced Q-factors, we have demonstrated the capability for pulsed resonant injection as well as the deterministic emission of coherent single photons from our devices. The anisotropic light-matter coupling in our elliptical micropillars, in combination with the principal capability for resonant injection and emission of indistinguishable photons puts our device platform in the forefront of engineering high-performance single photon sources. We foresee that carefully engineered devices and pump configurations will not only solve the problem of undesired photon loss associated with the commonly applied cross-polarization filtering, but ultimately lead to devices being capable of emitting linearly polarized single, coherent photons on demand with emission rates up to 10 GHz.

The authors would like to thank M. Emmerling and A. Wolf for sample preparation. We acknowledge financial support by the State of Bavaria and the German Ministry of Education and Research (BMBF) within the projects Q.com-H. Project HYPER-U-P-S has received funding from the QuantERA ERA-NET Cofund in Quantum Technologies implemented within the European Union's Horizon 2020 Programme. We further acknowledge funding by the DFG within the project SCHN1376-5.1 and PR1749/1-1. A. P. would like to acknowledge the support of the Swedish research council and the KAW foundation. T. L. was supported by the Ministry of Education (Singapore) grant 2017-T2-1-001. T. H. acknowledges support by the Elite Network Bavaria within the doctoral training programme "`Topological Insulators"' (Tols 836315).

\appendix*
\label{appendix}

\section{}

\subsubsection{Intensity of Linewidth of the polaritonic resonances}

\begin{figure*}[htb]
\begin{center}
\includegraphics[width=0.99\textwidth]{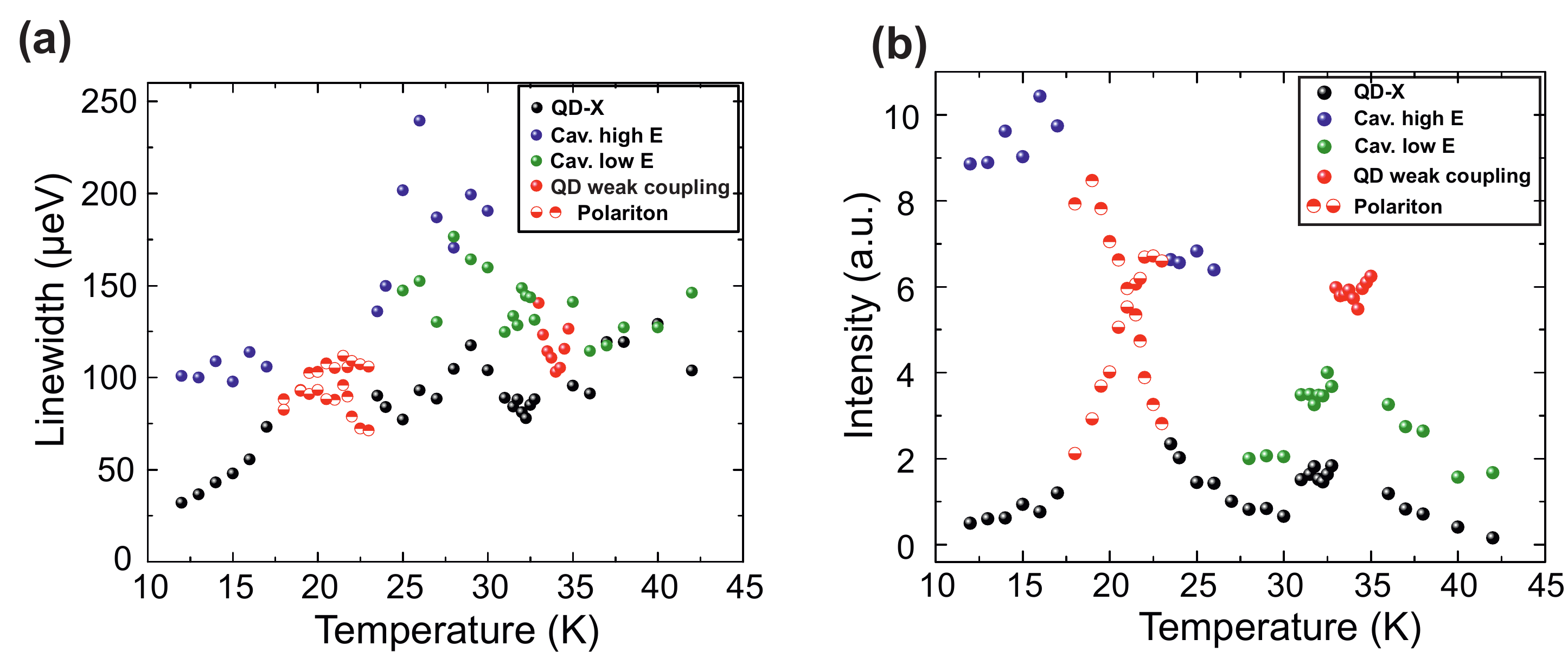}
\caption{ The linewidths as a function of the temperature are illustrated. At the anti-crossing one can observe a clear exchange of the linewidths of exciton and cavity. In the strong coupling regime the FWHM of both emission lines can be analysed, whereas the Purcell enhancement in the weak coupling regime enables only the analysis of the QD emission line. (b) depicts the exchange of the integrated intensity of the QD exciton and the cavity mode in the strong coupling regime, while in weak coupling only the quantum dot intensity was recordable. }
\label{fig:SC-linewidth}
\end{center}
\end{figure*}

In Fig.\ref{fig:SC-linewidth}(c) the linewidths as a function of the temperature are illustrated. Since the strong coupling regime provides a distinct splitting of the coupled modes in frequency domain, an extraction of both linewidths $\gamma_{C}$ and $\gamma_{X}$ for the whole temperature range is possible. Here, a clear exchange of the exciton and cavity linewidth can be observed\, \cite{Reithmaier2004}. 

Around temperatures of 32 K, we reach the weak coupling regime where only the QD's linewidth could be evaluated reliably. The increasing temperature raises the phonon influence which is expressed in an overall rise of the linewidth due to phonon-exciton interaction\, \cite{J.Iles-Smith2017, dusanowski2014phonon}. Similar to the exchange of the linewidths, the integrated intensities of QD and higher energy mode in Fig.\ref{fig:SC}(d) are exchanging at the anti-crossing, while in the weak coupling regime only the quantum dot intensity could be measured due to the Purcell effect.

\subsubsection{Theory of fine structure splitting in elliptical micropillars}

The longitudinal components of the electric field ($E_z$) and magnetic field ($H_z$) follow the wave equation:
\begin{equation}
\left(\frac{\partial^2}{\partial x^2}+\frac{\partial^2}{\partial y^2}\right)\left(\begin{array}{c}E_z\\H_z\end{array}\right)+\left(k^2-k_z^2\right)\left(\begin{array}{c}E_z\\H_z\end{array}\right)=0\label{eq:WaveEquation1}
\end{equation}
where $k_z=n\omega/c$ is the wavevector of the cavity mode in the growth direction, with $n$ the refractive index of the cavity and $\omega$ is its resonant frequency. It is convenient to introduce elliptical coordinates:
\begin{subequations}
\begin{align}
x&=\rho\cosh\xi\cos\eta\\
y&=\rho\sinh\xi\sin\eta
\end{align}
\end{subequations}
where $\rho=e_c a$, $e_c=\sqrt{a^2+b^2}/a$ is the eccentricity (not to be confused with ellipticity), and we recall that $a$ and $b$ denote the semi-major and semi-minor axes. In the elliptical coordinates, the boundary of the ellipse corresponds to $\xi=\xi_0=\cosh^{-1}(1/e_c)$ and the wave equation becomes [G. Blanch, p. 722 {\it in Handbook of Mathematical Functions} edited by M. Abramowitz \& I. A. Stegun, {\it Dover, New York} (1972)]:
\begin{equation}
\frac{1}{\rho'(\xi,\eta)}\left(\frac{\partial^2}{\partial\xi^2}+\frac{\partial^2}{\partial\eta^2}\right)\left(\begin{array}{c}E_z\\H_z\end{array}\right)+\left(k^2-k_z^2\right)\left(\begin{array}{c}E_z\\H_z\end{array}\right)=0\label{eq:WaveEquation2}
\end{equation}
where $\rho^\prime(\xi,\eta)=\frac{\rho^2}{2}\left(\cosh(2\xi)-\cos(2\eta)\right)$ is the differential area element in elliptical coordinates. 

From knowing the fields along the growth ($z$) direction, the transverse field components are given by:
\begin{subequations}
\begin{align}
E_{\eta}(\xi,\eta)&=\frac{i}{4q\sqrt{\sinh^2\xi+\sin^2\eta}}\left(k_z\frac{\partial E_{zi}}{\partial\eta}-\omega\mu\frac{\partial H_{zi}}{\partial\xi}\right)\\
E_{\xi}(\xi,\eta)&=\frac{i}{4q\sqrt{\sinh^2\xi+\sin^2\eta}}\left(k_z\frac{\partial E_{zi}}{\partial\eta}+\omega\mu\frac{\partial H_{zi}}{\partial\eta}\right)\\
H_{\eta}(\xi,\eta)&=\frac{i}{4q\sqrt{\sinh^2\xi+\sin^2\eta}}\left(k_z\frac{\partial H_{zi}}{\partial\eta}+\omega\epsilon\frac{\partial E_{zi}}{\partial\xi}\right)\\
H_{\xi}(\xi,\eta)&=\frac{i}{4q\sqrt{\sinh^2\xi+\sin^2\eta}}\left(k_z\frac{\partial H_{zi}}{\partial\eta}-\omega\epsilon\frac{\partial E_{zi}}{\partial\eta}\right)
\end{align}
\end{subequations}
where $\mu$ and $\epsilon$ are the relative permeability and relative permittivity, respectively, and $q=(k^2-k_z^2)\rho^2/4$.

The field solutions for $E_z$ and $H_z$ can be expanded in terms of Mathieu functions, which are composed of both angular and radial parts. The angular parts are given by the Mathieu cosine function, denoted $ce_n(\eta;q)$, and Mathieu sine function, denoted $se_n(\eta;q)$, respectively. For our problem, these functions must be periodic in $\eta$ with period an integer of $2\pi$. The index $n$ labels the number of nodes in the interval $0\leq\eta<\pi$. Inside the elliptical pillar the radial solutions are given by the modified/radial Mathieu functions of the first kind, which are divided into even functions, denoted $Ce_n(\xi;q)$, and odd functions, denoted $Se_n(\xi;q)$, where $n$ is an integer ($n\geq0$ for the even functions and $n\geq1$ for the odd functions). Outside the elliptical pillar boundary, we expect a decaying solution and this is given by the modified/radial solutions of the second kind. These are divided into even and odd functions (in a similar way to the first kind solutions). For $q>0$, the solutions are denoted $Fey_n(\eta;q)$ and $Gey_n(\eta;q)$, while for $q<0$, the solutions are denoted $Fek_n(\eta;q)$ and $Gek_n(\eta;q)$. Details on the calculation of the Mathieu functions and determination of the characteristic values are given in~[N. W. McLachlan, {\it Theory and Applications of Mathieu Functions}, Dover, New York (1964)]. We note that all the Mathieu functions can be conveniently expanded in series of Bessel functions, from which their derivatives can be also readily calculated.

For our problem, there are two relevant regions. Inside the micropillar, $\xi<\xi_0$, we have a refractive index $\sqrt{\epsilon_i}=n_i$, relative permeability $\mu_i=1$ (we assume a non-magnetic medium), and we will denote the electric and magnetic field solutions as $E_{zi}$ and $H_{zi}$, respectively. Outside the micropillar, $\xi>\xi_0$, we can take $\epsilon_o=1$ and $\mu_o=1$, and we will denote the electric and magnetic field solutions as $E_{zo}$ and $H_{zo}$, respectively. We can also define $q$ inside and outside the micropillar, respectively as:
\begin{subequations}
\begin{align}
q_i&=\frac{\frac{\omega^2n^2}{c^2}-k_z^2}{4}\rho^2\\
q_o&=\frac{\frac{\omega^2}{c^2}-k_z^2}{4}\rho^2
\end{align}
\end{subequations}
It is argued in Ref.~[K. Halterman \& P. L. Overfelt, Phys. Rev. A, {\bf 76}, 013834 (2007)] that the field solutions divide into two sets, where the first set has $H_z$ composed of even Mathieu functions and $E_z$ is composed of odd Mathieu functions, while the second set has $H_z$ composed of odd Mathieu functions and $E_z$ is composed of even Mathieu functions. Let us first consider the first case, where the field solutions inside the micropillar are:
\begin{subequations}
\begin{align}
E^e_{zi}(\xi,\eta)&=\sum_{m=1}^\infty a_{mi} Se_m(\xi;q_i)se_m(\eta;q_i)\\
H^e_{zi}(\xi,\eta)&=\sum_{m=0}^\infty b_{mi} Ce_m(\xi;q_i)ce_m(\eta;q_i),
\end{align}
\end{subequations}
where $\{a_{mi},b_{mi}\}$ are coefficients to be determined. Outside the micropillar, the solutions must decay, and are expanded:

\begin{subequations}
\begin{align}
E^e_{zo}(\xi,\eta)&=\sum_{m=1}^\infty a_{mo} Gek_m(\xi;-q_o)se_m(\eta;-q_o)\\
H^e_{zo}(\xi,\eta)&=\sum_{m=0}^\infty b_{mo} Fek_m(\xi;-q_o)ce_m(\eta;-q_o),
\end{align}
\end{subequations}

where $\{a_{mo},b_{mo}\}$ are coefficients to be determined.

To determine the allowed coefficients, the axial fields $E_z$ and $H_z$, as well as the tangential fields $E_\eta$ and $H_\eta$ must be continuous at the boundary $\xi=\xi_0$. Such conditions give rise to a hierarchy of equations, where the $\eta$ dependence should be integrated out. It is necessary to truncate the resulting infinite hierarchy of equations by limiting the number of terms in the solutions for the fields. For simplicity, we will consider only two terms in the field outside the micropillar.

It will be convenient to introduce the following notation:
\begin{subequations}
\begin{align}
Ce_m(\xi_0;q_i)&=c_m\\
Se_m(\xi_0;q_i)&=s_m\\
Gek_m(\xi_0;-q_0)&=g_m\\
Fek_m(\xi_0;-q_0)&=f_m\\
\left.\frac{dCe_m(\xi;q_i)}{d\xi}\right|_{\xi=\xi_0}&=c_m'\\
\left.\frac{dSe_m(\xi;q_i)}{d\xi}\right|_{\xi=\xi_0}&=s_m'\\
\left.\frac{dGek_m(\xi;-q_0)}{d\xi}\right|_{\xi=\xi_0}&=g_m'\\
\left.\frac{dFek_m(\xi;-q_0)}{d\xi}\right|_{\xi=\xi_0}&=f_m'
\end{align}
\end{subequations}

Then the axial fields at $\xi=\xi_0$ are:
\begin{subequations}
\begin{align}
E^e_{zi}(\xi_0,\eta)&=a_{1i}s_1se_1(\eta;q_i)\\
H^e_{zi}(\xi_0,\eta)&=b_{1i}c_1ce_1(\eta;q_i)\\
E^e_{zo}(\xi_0,\eta)&=a_{1o}g_1se_1(\eta;-q_o)+a_{3o}g_3se_3(\eta;-q_o)\\
H^e_{zo}(\xi_0,\eta)&=b_{1o}f_1ce_1(\eta;-q_o)+b_{3o}f_3ce_3(\eta;-q_o)
\end{align}
\end{subequations}

The tangential fields at $\xi=\xi_0$ are:
\begin{subequations}
\begin{align}
E^e_{\eta i}&(\xi_0,\eta)\notag\\
&=\frac{i\left(k_za_{1i}s_1se'_1(\eta;q_i)-\omega\mu_i b_{1i}c_1'ce_1(\eta;q_i)\right)}{4q_i\sqrt{\sinh^2\xi+\sin^2\eta}}\\
H^e_{\eta i}&(\xi_0,\eta)\notag\\
&=\frac{i\left(k_zb_{1i}c_1ce'_1(\eta;q_i)+\omega\epsilon_i a_{1i}s_1'se_1(\eta;q_i)\right)}{4q_i\sqrt{\sinh^2\xi+\sin^2\eta}}\\
E^e_{\eta o}&(\xi_0,\eta)\notag\\
&=\frac{ik_z\left(a_{1o}g_1se_1'(\eta;-q_o)+a_{3o}g_3se_3'(\eta;-q_o)\right)}{4q_o\sqrt{\sinh^2\xi+\sin^2\eta}}\notag\\
&-\frac{i\omega\left(b_{1o}f_1'ce_1(\eta;-q_o)+b_{3o}f_3'ce_3(\eta;-q_o)\right)}{4q_o\sqrt{\sinh^2\xi+\sin^2\eta}}\\
H^e_{\eta o}&(\xi_0,\eta)\notag\\
&=\frac{ik_z\left(b_{1o}f_1ce_1'(\eta;-q_o)+b_{3o}f_3ce_3'(\eta;-q_o)\right)}{4q_o\sqrt{\sinh^2\xi+\sin^2\eta}}\notag\\
&+\frac{i\omega\left(a_{1o}g_1'se_1(\eta;-q_o)+a_{3o}g_3'se_3(\eta;-q_o)\right)}{4q_o\sqrt{\sinh^2\xi+\sin^2\eta}}
\end{align}
\end{subequations}

First let us match the fields $E^e_{zi}(\xi_0,\eta)=E^e_{zo}(\xi_0,\eta)$:
\begin{equation}
a_{1i}s_1se_1(\eta;q_i)=a_{1o}g_1se_1(\eta;-q_o)+a_{3o}g_3se_3(\eta;-q_o)\label{eq:EzMatching}
\end{equation}
To remove the $\eta$ dependence, let us first define the following overlap integrals, which are the same as those given in Ref.~[D. A. Goldberg, L. J. Laslett, \& R. A. Rimmer, IEEE Transactions on Microwave Theory and Techniques, {\bf 38}, 1603 (1990)]:
\begin{align}
\alpha_{mn}&=\frac{1}{\pi}\int_0^{2\pi}ce_m(\eta;-q_o)ce_n(\eta;q_i)d\eta\\
\beta_{mn}&=\frac{1}{\pi}\int_0^{2\pi}se_m(\eta;-q_o)se_n(\eta;q_i)d\eta\\
\tau_{mn}&=\frac{1}{\pi}\int_0^{2\pi}se_m'(\eta;q_i)ce_n(\eta;-q_o)d\eta\\
\psi_{mn}&=\frac{1}{\pi}\int_0^{2\pi}ce_m'(\eta;q_i)se_n(\eta;-q_o)d\eta\\
\gamma_{mn}&=\frac{1}{\pi}\int_0^{2\pi}ce_m'(\eta;-q_o)se_n(\eta;-q_o)d\eta\\
\end{align}
It is important to note that any of the above overlap integrals vanish when $m$ is even and $n$ is odd or vice versa. Also, when $q_0=\pm q_i$, $\alpha=\delta_{nm}$ and $\beta=\delta_{nm}$.

Returning to Eq.~\ref{eq:EzMatching}, to remove the $\eta$ dependence, we can multiply by $ce_1(\eta;-q_o)$ or $se_3(\eta;-q_o)$ and integrate. This gives two equations:
\begin{align}
a_{1i}s_1\beta_{11}&=a_{1o}g_1\\
a_{1i}s_1\beta_{31}&=a_{3o}g_3
\end{align}
Similarly, from the matching of the fields $H^e_{zi}(\xi_0,\eta)=H^e_{zo}(\xi_0,\eta)$:
\begin{align}
b_{1i}c_1\alpha_{11}&=b_{1o}f_1\\
b_{1i}c_1\alpha_{31}&=b_{3o}f_3
\end{align}
We also obtain an equation from matching the fields $E^e_{\eta i}(\xi_0,\eta)=E^e_{\eta o}(\xi_0,\eta)$ (multiplying by $ce_1(\eta;-q_o)$ and integrating):
\begin{align}
&k_zq_oa_{1i}s_1\tau_{11}-\omega\mu_i q_o b_{1i} c_1'\alpha_{11}\notag\\
&\hspace{5mm}=-k_zq_ia_{1o}g_1\psi_{11}-k_zq_ig_3a_{3o}\gamma_{13}-\omega q_i b_{1o}f_1'
\end{align}
Finally, there is an equation from matching the fields $H^e_{\eta i}(\xi_0,\eta)=H^e_{\eta o}(\xi_0,\eta)$ (multiplying by $se_1(\eta;-q_o)$ and integrating):
\begin{align}
&k_zq_ob_{1i}c_1\psi_{11}+\omega\epsilon_i q_o a_{1i} s_1'\beta_{11}\notag\\
&\hspace{5mm}=k_zq_ib_{1o}f_1\gamma_{11}+k_zq_if_3b_{3o}\gamma_{31}+\omega q_i a_{1o}g_1'
\end{align}

\begin{widetext}

The field matching conditions can be neatly cast into matrix form:
\begin{equation}
\left(\begin{array}{cccc}s_1\beta_{11}&-g_1&0&0\\0&0&c_1\alpha_{11}&-f_1\\k_zs_1\left(q_o\tau_{11}+q_i\beta_{31}\gamma_{13}\right)&k_zq_ig_1\psi_{11}&-\omega\mu_i q_oc_1'\alpha_{11}&\omega q_if_1'\\ \omega\epsilon_iq_os_1'\beta_{11}&-\omega q_i g_1'&k_zc_1\left(q_o\psi_{11}-q_i\alpha_{31}\gamma_{31}\right)&k_zq_if_1\gamma_{11}\end{array}\right)\left(\begin{array}{c}a_{1i}\\a_{1o}\\b_{1i}\\b_{1o}\end{array}\right)=0
\end{equation}
The eigenvalues $\omega$ give the frequencies of the modes and the eigenvectors define the coefficients needed to define the mode in space. While the above treatment is for modes even in $H$, we can obtain the odd modes in $H$ by interchanging the even and odd functions: $c_m\leftrightarrow s_m$, $g_m\leftrightarrow f_m$, $c_m^\prime \leftrightarrow s_m^\prime$, $g_m^\prime \leftrightarrow f_m^\prime$, $\alpha_{mn}\leftrightarrow\beta_{mn}$, $\tau_{mn}\leftrightarrow\psi_{mn}$, $\gamma_{mn}\leftrightarrow\xi_{mn}=(1/\pi)\int_0^{2\pi}se_m'(\eta;-q_o)ce_n(\eta;-q_o)d\eta$.
\end{widetext}

\end{document}